# ICT's Effect on Parents' Feelings of Presence, Awareness, and Connectedness during a Child's Hospitalization

Abbas Heidari
Department of Computing and Information Systems
The University of Melbourne
Parkville, Australia
Email: abbasheidari@gmail.com

Yahya Kazemzadeh
Centre for Organisational and Social Informatics
Faculty of Information Technology
Monash University
Caulfield, Australia
Email: yahya.kazemzadeh@monash.edu

Greg Wadley
Department of Computing and Information Systems
The University of Melbourne
Parkville, Australia
Email: greg.wadley@unimelb.edu.au

## Abstract

This study evaluates how off-the-shelf commercial ICTs can contribute to creating a feeling of Presence, Connectedness, and Awareness between parents and their hospitalized child. Thematic analysis and descriptive statistics are used to analyse qualitative and quantitative data collected through a survey of thirty eight parents whose children were admitted to the Royal Children's Hospital in Melbourne, Victoria, Australia. Through analysis of data, Presence is found to be less facilitated through ICT than are Awareness and Connectedness. Although participants reported that voice call on mobile phones was the most common way of communication, their ideal was a video-chat application such as Skype, or a combination of Skype and TV to facilitate feeling of Presence. We discovered a strong desire by parents to use rich media such as video and audio to achieve a greater feeling of the Presence of their absent child.

## 1  Introduction

Hospitalization for children is an upsetting event (Thompson 1985). Hospitalized children are under stress, as hospital is not a familiar place for them compared to home and school (Bossert 1994). Isolation from familiar home and school environment impacts the emotional wellbeing of children receiving long term hospital care (Wadley et al. 2014). Regardless of the type of illness and whether it is acute or chronic, anxious children are likely to perceive hospitalization as a highly stressful experience (Bossert 1994). The stress created by hospitalization may result in physiological symptoms such as higher temperature, pulse rate and blood pressure as well as post-operative vomiting, disturbed sleep and an extended period of recovery (Skipper and Leonard 1968). Children with chronic illness are at higher risk of behavioural and emotional problems and psychiatric disorder (Hysing et al. 2007).

The effects of hospitalization are not limited to the children themselves. Families of hospitalized children also experience challenges such as feelings of isolation, fear, and anxiety regarding their child's health. Increasing stress due to illness and hospitalization, feelings of guilt as the result of being far away, are other effects of hospitalization. Physical and geographical separation may result in difficulties in the parents' relationship (Nicholas et al. 2011).

Communication via Information and Communications Technology (ICT) has been shown to reduce stress in hospitalized children (Nicholas et al. 2011). Researchers who target the hospital domain have designed and trialled novel technologies for hospitalized children to connect to their families (Hopkins et al. 2014; Nicholas et al. 2011; Parsapour et al. 2011) and school (Green et al. 2011). Recently a variety of new commercial ICTs have become available that might be useful in this context. However, little is known about whether and how these technologies are being used by hospitalized children and their parents, nor their preferences for technology or devices. In particular, the impact of mobile smart phones and tablets has not been studied. To study how families are deploying off-the-shelf technology to maintain connections with their children in hospital, we used surveys to collect information from parents at the Royal Children's Hospital (RCH), a public hospital in Melbourne, Australia.

Communication researchers have studied mediated social connection and found that a sense of Connectedness with a distant other is associated with Awareness of their activities and a sense of the their Presence. We analysed our data to discover how these phenomena are involved in media use by families in the hospital context.

## 2 Prior Work

ICTs have been shown to be able to strengthen family bonds, expand psychological neighbourhoods, and facilitate social relationships (Wei and Lo 2006). Interpersonal ties can be maintained by mediated connections via ICT (Wellman and Gulia 1999). Social Connectedness is negatively correlated with anxiety, loneliness, social distress avoidance, social discomfort, and a host of other negative emotions (Baumeister and Leary 1995; Lee and Robbins 1998; Lee et al. 2001). Therefore, understanding how ICT can create Connectedness might help in solving problems of isolation that hospitalized children face.

Therefore, we begin by reviewing the communication concepts Presence, Awareness, Connectedness, and their relationships. We then review prior research on the use of ICT in hospital settings. We provide an assessment of the uptake of off-the-shelf ICTs by Australian families. Finally, our research gaps and research questions are presented.

Presence has been defined as a technology user's subjective sensation of "being there" in a scene represented by a medium (Barfield et al. 1995). Presence is the subjective feeling of existing within a given environment (Heeter 1992; Sheridan 1992). Slater and Usoh (1994) have defined Presence as "the (suspension of dis-) belief" in a world which is different from the real world. Presence concerns the degree to which a medium can create a representation that looks, sounds and feels like the "real" thing from accurate representation of objects, event and people (Lombard and Ditton 1997).

"Copresence" has been defined as "being there together" (Schroeder 2006). Copresence is sometimes called Social Presence (Biocca and Harms 2002; Rice 1992; Short et al. 1976). It refers to a sense of being together with others via a communication medium or virtual environment.

Early communication research found that visual media conveys more Social Presence than verbal (e.g. audio) media, which in turn conveys more Social Presence than written media (e.g. text) (Short et al. 1976). To generate Presence the number of sensory output channels is important as well as the consistency of information in different modalities because the information from all channels must describe the same objective world (Lombard and Ditton 1997). Copresence deals with the human-human relations, while Presence describes human-object relations (Zhao, 2003). In this paper we use the term "Presence" to mean "Social Presence".

The state of knowing the environment, presence, and activities of others has been defined as Awareness (Rowan and Mynatt 2005). Keeping the channels open for individuals to share their experiences with others, and maintaining human relationships are the aims of Awareness systems (Ijsselsteijn et al. 2003; Vetere et al. 2009). Dourish and Bellotti (1992) have defined Awareness as "an understanding of the activities of others, which provides a context for your own activity". It is an external perception either synchronous or asynchronous (Rettie 2003). For example finding a gift tax invoice from a loved one before receiving the gift will create Awareness and can create a reaction for that act.

Forty one Awareness phrases have been classified which can be categorized into four categories of Awareness (Christiansen and Maglaughlin 2003). These four categories were Workplace Awareness, Availability Awareness, Group Awareness and Contextual Awareness. Workplace Awareness is about Awareness of tasks within virtual environment while Availability Awareness is associated with people and objects. 'Feelings of belonging to a group' phrases have been associated with Group Awareness. In Contextual Awareness, physical, social and mental contexts have been included (Christiansen and Maglaughlin 2003).

More recently, researchers have shown that Awareness systems could help family members who were geographically distributed to feel connected to other relatives (Greenberg et al.; 2008, Metaxas et al.;

2007). They also felt safe or reassured (Brown et al. 2007; Metaxas et al. 2007; Mynatt et al. 2001), and more involved in each other's lives (Khan et al. 2007; Markopoulos et al. 2004). Another study identified the advantages for hospitalized children in having awareness of activities in the distant location (Wadley et al. 2014).

Connectedness has been defined as a positive emotional appraisement due to an ongoing social relationship that facilitates "staying in touch" (Ijsselsteijn et al. 2003). Furthermore, it has been recognized as a need in psychology (Rettie 2003). Feeling in touch or having a sense of Awareness can be created via ICT (Rettie, 2003).

## 2.1 Relationships between Presence, Awareness, and Connectedness

Awareness and Connectedness have been found to be highly correlated. Dey and Guzman (2006) found that when Awareness happens continuously, it facilitates learning about the daily routines of a friend or a loved one, which can lead to a sense of Connectedness. Connectedness is linked to Social Presence (Biocca et al. 2001) from a psychological engagement aspect (Rettie 2003). However, they are not the same. For instance, a simple "goodnight" text message can bring the emotional experience of Awareness, whereas status updates in online instant messaging (IM) transmits Awareness even when there are no message exchanges (Rettie 2003).

Awareness can be created without Connectedness or Social Presence. For example it is possible to have Awareness of what an employee is doing at work without having the emotional feeling of Connectedness or being there. Social Presence can happen with or without Connectedness. For instance, Social Presence in telemarketing can happen without emotional feelings of Awareness. However, Social Presence will enhance Awareness between two users.

Connectedness can be created via communication as a feeling of being in touch (Rettie 2003). Townsend (2002) suggests ICT can create Connectedness as mobile phones are enabling us to feel connected and thereby prevent loneliness. The effects of ICT on creating Connectedness, Awareness and Presence have been studied before. Dey and Guzman (2006) designed physical, peripheral Awareness displays named Presence Displays which illustrated the online presence of a close friend or family. Nine participants in this study experienced significant increase in level of Awareness and Connectedness by using Presence Display.

Townsend (2002) suggests ICT can create Connectedness as mobile phones are enabling us to feel connected, and they prevent loneliness. The effects of ICT on creating Connectedness, Awareness, and Presence have been studied before. Dey and Guzman (2006) designed physical, peripheral Awareness displays named Presence Displays which illustrated the online presence of a close friend or family. Nine participants in this study experienced significant increase in level of Awareness, and Connectedness, by using Presence Display (Dey and Guzman, 2006).

## 2.2 Mediated Connection for the Hospital Context

Green et al. (2011) found that ICT had positive impact on child's feeling of Presence, Awareness, and Connectedness at hospital and their classmates (Green et al. 2011) Parsapour et al. (2011) investigated the effects of video conferencing to maintain close contact between patients, family and friends during hospitalization. The results demonstrated that video conferencing can overcome some obstacles that limit family participation.

Bensink et al. (2006) examined the use of a videophone with an eight-year-old hospitalized boy who was diagnosed with attention deficit hyperactivity disorder (ADHD), and received a bone marrow transplant. The boy had several aggressive outbursts during his pre-transplantation hospitalization. Results of this study indicated the videophone significantly reduced the anxiety and the distress experienced by both the family and the child (Bensink et al. 2006).

## 2.3 Uptake of ICT by Australian Families

According to the Australian Bureau of Statistics (Australian Bureau of Statistics 2012b), just 1.7% of Australians accessed the internet in 2006 via mobile and fixed wireless. This proportion increased dramatically to 47.8% in 2014. Advances in technology enhanced people including parents to have access to higher technology at more affordable prices.

This technology has changed the way people communicate with each other. For instance in 2006, 65% of children aged 5 to 14 years accessed the internet, while in 2012 90% of them accessed the internet (Australian Bureau of Statistics 2012a). According to recent statistics, 29% of children aged 5 to 14 have mobile phones.

Therefore families with children in hospital are well-placed to exploit off-the-shelf broadband technology to promote connectedness among family members.

## 2.4 Research Gap

Considering consumer's vast usage of technology (Australian Bureau of Statistics 2012a), the positive results of using ICT to resolve some problems at hospital (Parsapour et al. 2011) and positive effect of presence of family members on ICU patients (Morgan et al. 2005), we felt it is important to study how families are currently using ICTs to facilitate Presence, Awareness and Connectedness between parent and their child at hospital via ICT to resolve some problems needed to be investigated.

Whereas prior research designed and evaluated bespoke technology, how families adopt existing commercial ICTs to stay connected with their hospitalized children is not well understood. Previous research identified the effects of customized devices or technologies to create Presence (Hopkins et al. 2014), Awareness, and Connectedness. However, none of these experimental technologies are in widespread use by parents. It is likely that parents and their children are discovering solutions to this problem themselves by using commercially-available off-the-shelf technologies. However, it is not known what parents are using at the current stage, and what difficulties they are facing with. This knowledge can help researchers to design and create technologies to be used by a wider range of users.

Advances in communication technology have changed the way people live and work. The studies done before were based on available technologies at the time of study. A recent study requires investigating how parents are using current ICT technology to communicate with their child at hospital.

## 2.5 Research Questions

To address the identified research gap, we devise the following research question (RQ), as well as three sub-questions. The major question we attempt to address is:

**Main RQ:** *How do off-the-shelf ICTs impact parents' feelings of Awareness, Connectedness, and Presence with respect to their hospitalized children?*

We address this via three sub-questions:

- RQ1: *What technologies are parents using to communicate with their child in hospital?*
- RQ2: *Do existing technologies help parents to achieve Presence of, Awareness of, and Connectedness with their child?*
- RQ3: *What difficulties do parents experience when using these technologies to communicate with their child?*

## 3 Approach

Field research involving observation with ill children or at the home of their parents was impossible, as was applying experimental research. Surveys are suitable for research questions about self-reported beliefs or behaviour (Neuman 2005). Therefore, a survey was selected as the method of data collection. Hospital settings are very stressful, especially for the parents of ill children; taking time away from their child might not be feasible. Therefore questionnaire was chosen.

In the first section of the questionnaire, demographic information of both parents and children was collected. Then we used five-point Likert scales to evaluate the knowledge of parents about their child at the hospital when they are not with them. In the third section of the questionnaire, we asked whether parents use technology to communicate with their child.

Then three open ended questions were asked. The first question was designed to examine the existence of patterns which evaluate the sense of Awareness, Connectedness, and Presence in parent-child conversations. Asking parents about their problems in using technology was useful to identify problems they have when they use a specific technology. Discovering the ideal technology or device that parents would like to use to communicate with their child was the inspiration to design the third question in this section. The gap between problems of current technology and the potential of ideal technology can help researchers to create a more useful device or technology to be used by widespread use of parents.

Thirty eight parents of children admitted to the RCH participated in this study. At the time of the survey, they were present at the hospital. We visited "Inpatient wards" including Surgical care, Surgical and Neuro care, Cardiac surgery, Intensive care, Surgical short stay, Cancer care, Medical care, Adolescent and Rehabilitation care during this time. As well as "Inpatient wards", "Outpatient wards" such as Day Medical and Specialist clinics were visited for collecting data. Table 1 demonstrates demographic information of participants. Thematic analysis was used to analyses qualitative data to search for themes emerged from data to help describing the phenomenon (Daly et al. 1997).

| Characteristic | Number of participants |
| --- | --- |
| Age | |
|     5–10 years | 15 |
|     11–15 years | 23 |
| Gender | |
|     Boys | 22 |
|     Girls | 16 |
| Child has been previously admitted to hospital | |
|     Yes | 7 |
|     No | 30 |
|     No reply | 1 |
| Length of hospital stay this time | |
|     <5 days | 15 |
|     >5 days | 19 |
|     No reply | 4 |
| Frequently of visiting child at hospital | |
|     Always (e.g. every day) | 35 |
|     Often (e.g. every second day) | 1 |
|     Sometimes (e.g. a few times per week) | 0 |
|     Rarely (e.g. less than once a week) | 0 |
|     No reply | 2 |
| Travel time from home to hospital | |
|     Shortest distance | 15 Minutes |
|     Longest distance | 260 Minutes |
|     Average distance | Mean 80.7 (SD 66.6) Minutes |

*Table 1: Background characteristics of participants (n = 38)*

## 4   Results

As shown in figure 1, parents indicated high satisfaction about their Awareness of their child's activities. The same applied to Connectedness to their child's activity. Also, parents claimed that they were mostly aware of their child's emotional feelings such as being happy, sad, or bored. However, they were not satisfied at the same degree about this Awareness.

Twenty-nine out of thirty-eight parents used technology to communicate with their child in the hospital. As demonstrated in Figure 2, among all different types of devices, "Mobile Phone" was the most popular device for their child to communicate with their parents. "Laptop" is the least used device for children to communicate with their parents. We found that using audio technology for making phone calls is the most popular medium.

As demonstrated in Figure 3, Presence was least likely to be created by communication devices, while Connectedness was the most, and Awareness was in the middle. As audio communication is a leaner medium compared to audio-visual communication, it is less able to create feeling of Presence (Lombard and Ditton 1997). On the other hand, since video was not used by twenty nine of the parents for communication, it was hard for them to have feeling of Presence.

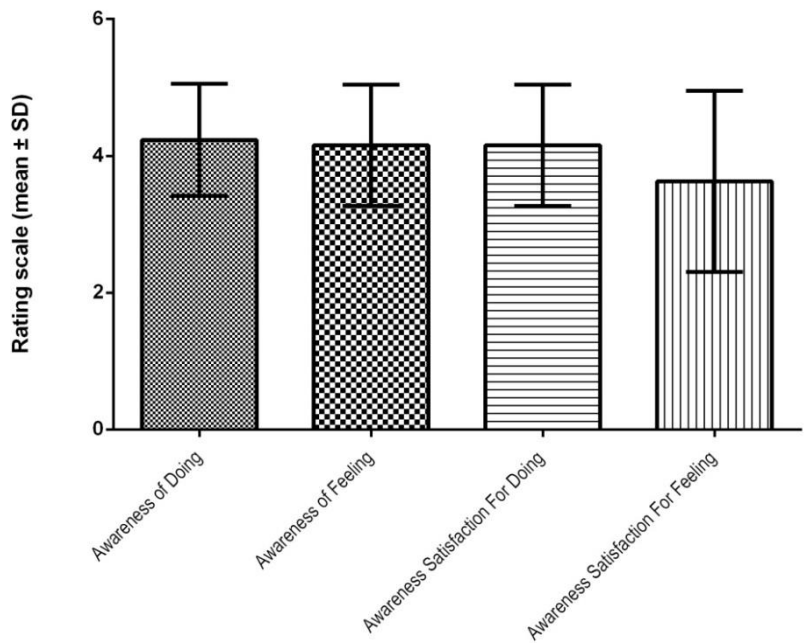

*Figure 1: Parental knowledge of their child circumstances and their satisfaction with this*

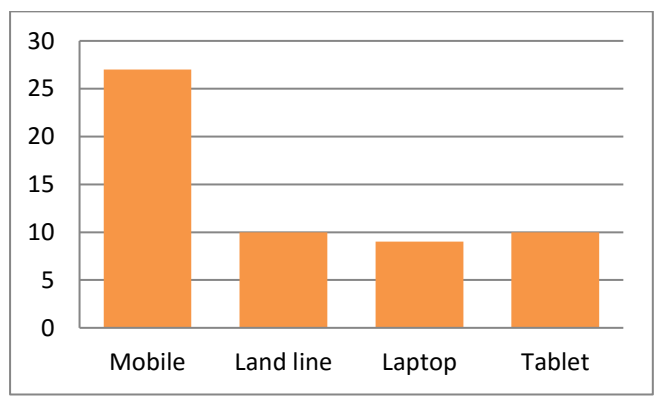

*Figure 2: The devices used in patent-child communication*

Parents were asked to explain the topics they usually talk with their child during mediated conversation. Thematic analysis resulted in seven categories of topics: "wellbeing", "news"," child's daily activities" , "health concerns", "things to bring to hospital" and "future plans".

Discovering difficulties in parent-child communication can create insights into what needs to be overcome to enhance mediated communications. We discovered four categories including "technical problems", "bad reception", "technology is not suitable" and "No difficulties with the technology".

Finally, we asked parents what would be the ideal technology to communicate with their hospitalized child. Responses included: "Visual communication", "Skype on TV", "24/7 in touch"," A mobile device", "Time machine"," Existing technologies are good" and "Physical contact at a  distance".

## 5  Discussion

The aim of this study was to find out how parents use off-the-shelf ICTs to communicate with hospitalized children, and how these technologies impact parents' feelings of Awareness, Connectedness, and Presence. We explored this question via three sub- questions:

RQ1: *What technologies are parents using to communicate with their child in hospital?*

The use of communication technology by twenty nine out of thirty eight parents demonstrates the popularity of ICT in the hospital context. Since parents of hospitalized children can be extremely busy maintaining family life while their child is in hospital, we might assume that technology use might reduce the need to make daily visits to the hospital. However, our research shows that almost all parents still attend hospital almost every day. Therefore, current technology is not able to substitute for visits satisfactorily. Either *the available technologies cannot satisfy their needs,* or *the way they use technologies is not optimum*. Parents' use of devices did not help them to feel the Presence of their child, though it more effectively delivered Awareness and Connectedness. *We conclude that, if technology can create the feeling of Presence, we can see a dramatic increase in parents' satisfaction.*

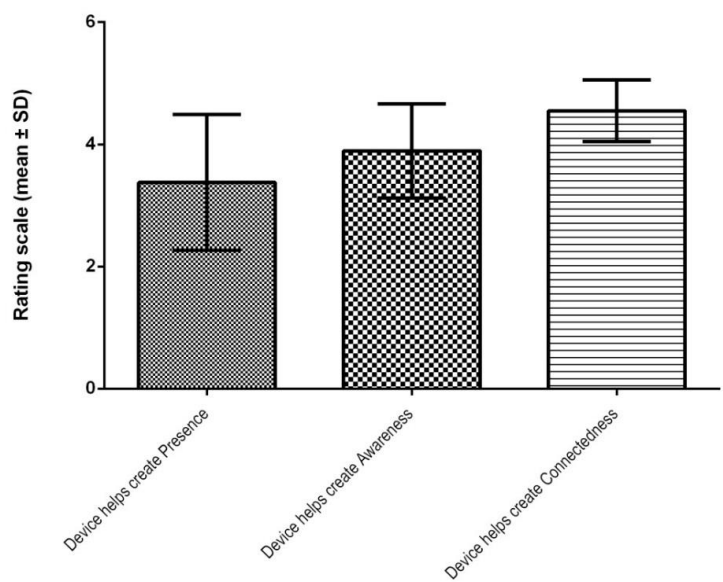

*Figure 3: Presence, Awareness and Connectedness created by using a device by parent*

As technology has advanced, expectations of parents in parent-child communication settings has moved from Connectedness and Awareness to a desire to experience the Presence of their child. We found that audio is the most used technology by parents and their children. However, parents wanted to have visual communication, such as Skype, in order to feel Presence. Prior literature asserts that the greater the number of human senses that a medium provides for simulation, the greater that medium produces a sense of Presence (Anderson and Casey 1997; Barfield and Weghorst 1993; Kim 1996; Short et al. 1976). Also, Lombard and Ditton (1997) state that media which provide both audio and video can create a greater sense of Presence compared to audio - or video - only media. Our findings confirm these prior findings.

A hospitalized child's illness may limit their movement or their ability to hold objects in their hands. For instance, they might not be able to sit in front of a PC or use a laptop for a long period. Our research demonstrates that they tend to use a device which is easy to use for their communication, such as a mobile phone. This is consistent with Fels et al. (2003) who found that ease of use can play an important role in device adoption among hospitalized children.

RQ2: *Do existing technologies help parents to achieve Presence of, Awareness of, and Connectedness with their child?*

We found that current technologies more successfully create Connectedness and Awareness but are less successful at creating a sense of Presence. By analysing qualitative and quantitative results, we found that the feeling of Awareness was most easily created via using a communication device. Conversations between parents and their child helped parents to be aware of news, their health conditions, how they are, what activities they have done, and would do in future, even activities that were happening around them like visitors or teachers' activities. This helped them to be aware of their child's environment, and the activities of others surrounding their child. The child was also able to gain Awareness about family, home, and pets in the world outside the hospital.

Although parents were satisfied about their level of Awareness, the desire for a "24/7 in touch" technology demonstrates the need to have more. Awareness can lead Connectedness (Dey and Guzman 2006). Twenty nine out of thirty eight parents used technology to communicate with their child; it is clear from their conversation topics that they were in consistent information exchange. Interviews were consistent with quantitative information gathered in our survey. Parents acknowledged that using their device helped them to feel connected with their child.

Some parents' answers in regards to communication problems fitted into the "Technology is not suitable" category. Within this group the interest in feeling Presence can be clearly seen when they complained that *"we can gauge much from Josh's body language & facial expressions than by his words only, so we find that difficult when we can't see him & can only hear his voice"* Or *"No technology device will replace the tactileness of touching feely"*. The need for a medium to give them the feeling of *"being there"* or *"being there together"* to be able to see or touch their child can be abstracted from their answers.

In thematic analysis of parents' expression of ideal technology, patterns related to Presence emerged more than Connectedness and Awareness. *"Visual communication"*, *"Skype on TV"* and *"Physical contact at a distance"* were related to Presence. The request for a video-audio communication such as *"Visual communication" and "Skype on TV" is* very clear. This demonstrates their strong desire for having more feeling of Presence. *"Physical contact at a distance"* shows the need to have the feeling of being there to be able to touch.

RQ3: *What difficulties do parents experience when using these technologies to communicate with their child?*

We discovered three groups. The first wanted to use technology to communicate but had difficulties in using it. For example, they experienced *"bad mobile reception"*. The second group didn't find technology useful for connecting with their child. For instance, one parent said *"No technology device will replace the tactfulness of touching feely* [sic]*"*. The third group mentioned they had no difficulties communicating with their child at the hospital.

Among the first and second group of parents' answers a theme of strong desire to have a rich mediated communication is apparent, although just nine parents used video for their communication. One parent described the problem of: *"Not being hooked up to Skype to see them & surroundings etc."*. Another parent said *"Skype tending to drop in & out with signal which is so frustrating…"*. Desiring to have a rich medium such as video can be seen in another parent's comment in the second group as he/she said *"I can gauge much from Josh's body language & facial expressions than by his words only, so I find that difficult when I can't see him& can only hear his voice"*. It appears that audio-visual communication is a standard to compare other media to. Nine parents in this study used video to communicate with their child. It is interesting to see there were no demands for text-based communication such as Facebook.

Based on our answers to RQ1, RQ2 and RQ3, we conclude that ICT can impact the feelings of Connectedness and Awareness in hospital settings. However, a rich audio-visual medium is required to facilitate a sense of Presence. In addition, parents are less satisfied about the Awareness of what their child is feeling than the Awareness of what their child is doing. The reason for this may be that they may already know or can predict the feelings of their child from their child's body language; as one parent mentioned *"I can gauge much from Josh's body language & facial expressions than by his words only"*.

Based on five premises, we concluded that there has been a move from being satisfied with having Awareness and Connectedness to a strong desire for Presence. First, as ICT has improved, usage of ICT has increased too. Second, previous research demonstrates that ICT can create Awareness, Connectedness, and Presence in hospital settings (Green et al. 2011). So, we have many users using ICT that can help them have Awareness, Connectedness, and Presence. Third, we found that parents' use of ICT helps them to have Awareness and Connectedness more than Presence. Fourth, thirty five out of thirty eight parents visited their child on daily basis. Fifth, when parents were asked to describe their ideal technology, a strong theme of video-audio mediated communication emerges from their answers. Considering the finding of Lombard and Ditton (1997) that an audio-visual communication is more successful at creating Presence compared to just video or audio communication, parents ask for having more Presence. Therefore, we conclude that, there is a move from being satisfied with having Awareness and Connectedness to a strong desire for Presence. Finally, we discovered from parents' answers to questionnaire that "Skype on T.V." is commonly stated as an ideal technology/device description.

## 6  Conclusion

Research to evaluate use off-the-shelf technology in hospital-family setting by parents and their child has not been conducted before. Our study found that families mostly use mobile phones for their communication. However, they strongly desired to have audio-visual communication to enhance their feeling of Presence. Prior research has examined the ability of specialized ICT to facilitate Awareness, Connectedness, and Presence in hospital-school settings (Green et al. 2011). Our study extends this, and demonstrates that off-the-shelf ICT can also facilitate Awareness and Connectedness, and to a lesser extent Presence, in parent-child relationships.

This study supports findings of Eröz-Tuğa and Sadler's (2009) who found that between six computer-mediated video chat tools, Skype and MSN Messenger were preferred and trusted tools for video chat. We found that Skype is parents' favourite application for video chat. Current technology has been successful in facilitating feelings of Awareness and Connectedness. Technology has improved and user expectations of an ideal communication technology now include a combination of audio and video which can enhance their feeling of Presence. When parents can have the feeling of being in the hospital with their child via a device, it may reduce travel costs.

Until we get significant improvement in facilitating the sense of Presence in parent-child hospital relationships, parents must suffer the expenses involved in traveling, which often involves time away from work and other family commitments. Technology cannot replace daily visits of parents completely, as there may be some psychological needs for face to face visits. However, it can decrease the number of visits to the hospital.

Many of the problem of using technology for communication with hospitalize children are technical. Some of these have simple solutions in practice: for instance, lack of Internet access at hospital can be resolved by obtaining broadband and sharing it with patients. Thereby, families could use their own device to have good quality communication, and more chance to have feeling of Presence.

## 7  Limitations and Future Research Directions

There were limitations for data collection at RCH related to time and resources. A more in-depth data collection method such as one-to-one interview may well create new knowledge in this area. Some parents in this study were selected from different long stay wards at RCH. However, some of them were approached when they attended at "day medical" or "special clinic" to see a doctor or specialist. Therefore, their stay at hospital was short and they were with the child all the time during the visit. There were difficulties with collecting data mainly from long term patients at RCH.

Future research should explore whether Presence may be better mediated if made via audio-visual communication. For this research, tablets could be handed out to parents and their child in pairs. One tablet can be passed to a parent and other one to the child in the hospital, and a video-audio application such as Skype used. The impact of audio-visual communication in facilitating Presence could then be studied. It can be broadened to extended family members such as uncles and aunts to discover how they acquire a feeling of Presence with the child in hospital. The effect of parents' presence has been shown to have positive physiological effects on ICU patients (Morgan et al. 2005). However, future research requires an evaluation of a possible positive effect of Presence on chronically ill children from a physiological perspective.

## Acknowledgements

The authors would like to acknowledge Frank Vetere for supervising this research. We also thank families for participating at a time of stress and uncertainty.

## Copyright